\documentclass[a4paper,11pt]{article}
\usepackage{pos}
\usepackage{booktabs}

\title{Fitting fractions of the $X_{\rm max}$ distributions at ultra high energies}
 \ShortTitle{Mass composition of UHECRs}

\author*[a]{Nicusor Arsene}

\affiliation[a]{Institute for Space Sciences, \\
P.O.Box MG-23, Ro 077125, Bucharest-Magurele, Romania}

\emailAdd{nicusorarsene@spacescience.ro}

\abstract{
The mass composition of ultra high-energy cosmic ray (UHECRs) can be inferred from measurements of $X_{\rm max}$ distributions by fitting them with Monte Carlo (MC) predictions for different primary species of nuclei in each energy interval. On the basis of Monte Carlo (MC) simulations, we show that an appropriate approach is to fit the observed $X_{\rm max}$ distributions with all possible combinations of elements from a large set of primaries (in our case p, He, C, N, O, Ne, Si and Fe), and to find the "best combination" of elements which best describe the observed $X_{\rm max}$ distributions. 
We apply this method to the $X_{\rm max}$ distributions recorded by the Pierre Auger (2014) and Telescope Array (TA) (2016) Observatories in the energy range $\lg E (\rm eV) =$ [17.8 - 19.3] and 
$\lg E (\rm eV) =$ [18.2 - 19.0], respectively, by employing MC predictions of the QGSJETII-04 hadronic interaction model. The results obtained from both data sets suggest that the mass composition of UHECRs is dominated by protons and He nuclei ($\gtrsim 70\%$) which present a modulation of their abundances as a function of primary energy, but keeping their sum roughly constant. We performed an indirect comparison between the two data sets measured by the two experiments and found a good degree of compatibility in some energy bins around and above the \textit{ankle} ($\lg E (\rm eV) \sim 18.7$), but worsening at lower energies.
We consider that the current approach, completed with predictions of different hadronic interaction models, can be used in further studies on mass composition to obtain a more accurate image of the evolution of the individual fractions of nuclei as a function of energy on the basis of experimental $X_{\rm max}$ distributions. 
}

\FullConference{37$^{\rm{th}}$ International Cosmic Ray Conference (ICRC 2021)\\
		July 12th -- 23rd, 2021\\
		Online -- Berlin, Germany}

\begin{document}
\maketitle

\section{Introduction}

For a better understanding of the origin and acceleration mechanisms of UHECRs it is absolutely necessary a very accurate reconstruction of the mass composition along with their energy spectrum and arrival directions using the ground based cosmic ray experiments such as the Pierre Auger \cite{ThePierreAuger:2015rma} and Telescope Array (TA) \cite{Tokuno:2012mi} observatories. The mass composition is reconstructed on the basis of the $X_{\rm max}$ parameter, the atmospheric depth where energy deposit profile of secondary particles from extensive air shower (EAS) reaches its maximum. This parameter is related to the mass of the primary particle as $\langle X_{\rm max}\rangle \propto - \ln A$.
Measurements of $\langle X_{\rm max}\rangle$ and $\sigma(X_{\rm max})$, the first two moments of an $X_{\rm max}$ distribution, were used to infer mass composition as a function of the primary energy in different experiments \cite{ThePierreAuger:2015rma, PhysRevLett.104.091101, Aab:2014kda, PhysRevLett.104.161101, Abbasi:2018nun}. 
However, it was found that using only the limited information given by the first two moments of the $X_{\rm max}$ distributions may result in some degeneracies when interpreting the mass composition, since different combinations of primary elements can reproduce exactly the same mean and dispersion \cite{Aab:2014aea}. 
In the same paper was proposed a method which uses the information of the full shape of $X_{\rm max}$ distributions, by fitting them with Monte Carlo (MC) templates for four fixed primary species (p, He, N and Fe) on the entire energy spectrum. Following this approach, information about the evolution of the abundances of individual nuclei as a function of primary energy was obtained.

Different astrophysical models suggest variations of the abundances of different primary species as a function of energy related to the acceleration and propagation scenarios. If such variations would really exist in nature, we consider that fitting the $X_{\rm max}$ distributions with the same four elements on the entire energy spectrum, the reconstructed fractions of the individual nuclei will be biased in some energy intervals as a consequence of not including into the fitting procedure of some intermediate elements which are in fact present.
On the basis of MC simulations we show that fitting the $X_{\rm max}$ distributions with the four fixed elements (p, He, N and Fe), the quality of the fit is affected if some intermediate elements, e.g. Ne/Si are in fact present in a quite large abundance ($> 40 \%$) \cite{Arsene:2020ago}. We present an alternative approach which fits the $X_{\rm max}$ distributions with all possible combinations of elements from a larger set of primaries (p, He, C, N, O, Ne, Si and Fe), finding in this way the "best combination" of elements which best describe the observed $X_{\rm max}$ distribution. We apply this method to the $X_{\rm max}$ distributions recorded by the Pierre Auger (2014) and TA (2016) using MC predictions computed with CONEX v4r37 \cite{Pierog:2004re,Bergmann:2006yz} employing the QGSJETII-04 \cite{Ostapchenko:2005nj} hadronic interaction model. Finally, we perform an indirect comparison between $X_{\rm max}$ distributions recorded by the two experiments, on the basis of the individual fractions of nuclei reconstructed with this procedure, using MC predictions of QGSJETII-04 hadronic interaction model.

\section{MC templates}

\begin{figure}
\centering
    \includegraphics[width=0.8\textwidth]{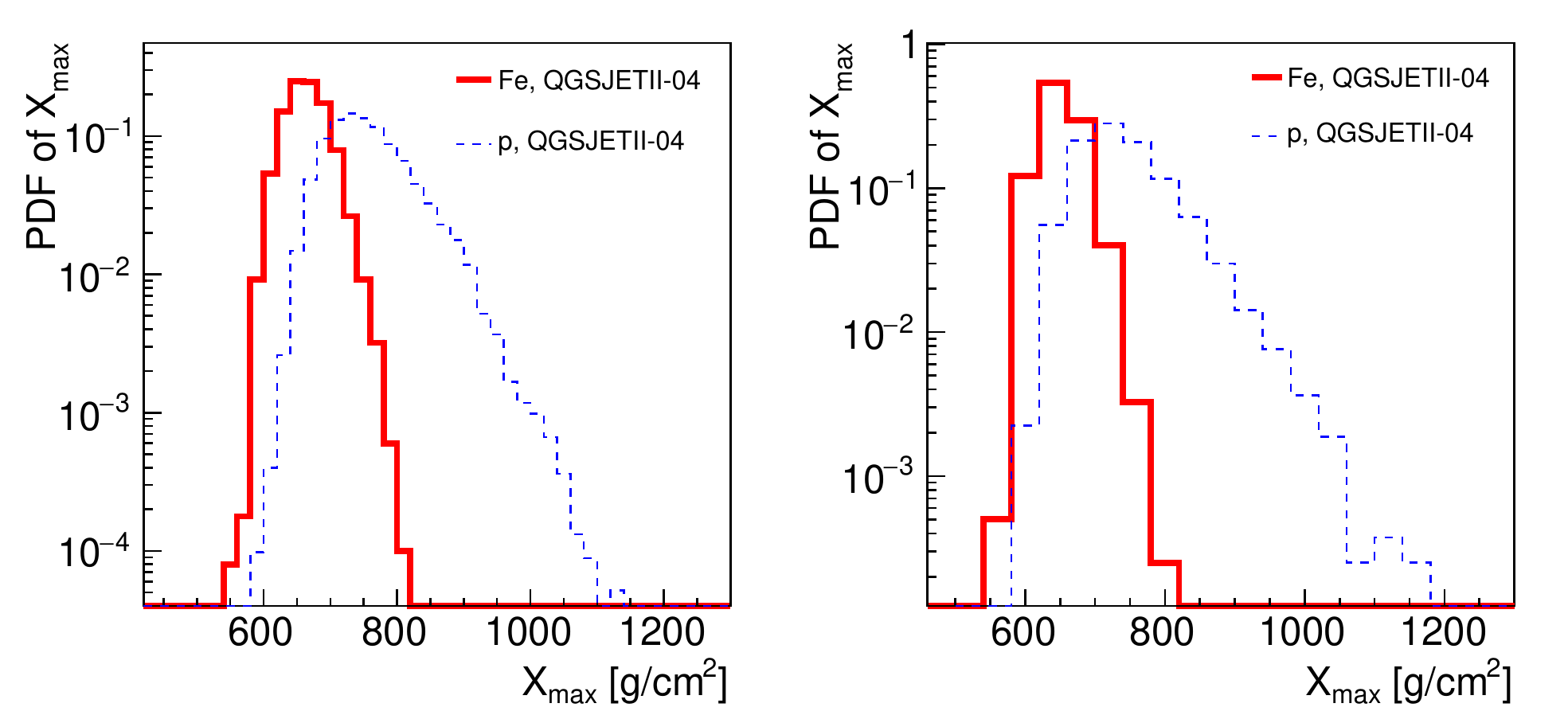}
    \caption{PDFs of $X_{\rm max}$ for proton and iron induced showers in the energy range lg ($E$/eV) $= [18.4 - 18.5]$ employing the QGSJETII-04 hadronic interaction model for Auger case \textit{(left)} and TA case \textit{(right)}.}
    \label{fig_pdfs}
\end{figure}

The MC templates were computed with  CONEX v4r37 for each primary species (p, He, C, N, O, Ne, Si and Fe) in each energy intervals of $0.1$ in lg ($E$/eV)  between lg ($E$/eV) = [17.8 - 19.3] and lg ($E$/eV) $= [18.2 - 19.0]$ for Auger and TA case, respectively. A PDF of $X_{\rm max}$ consists in a binned 1D histogram in the range $[0 - 2000]$ g/cm$^{2}$ with  a bin width of $20$ g/cm$^{2}$ for Auger and $[500 - 1300]$ g/cm$^{2}$ with the bin width = $40$ g/cm$^{2}$ for TA. Each true $X_{\rm max}$ value computed by CONEX filling the PDFs are modified in accord with the acceptance and resolution of each experiment. For the case of Auger, the true $X_{\rm max}$ values were modified by making use of the Equations (7) and (8) from \cite{Aab:2014kda}, while for the TA case the true $X_{\rm max}$ values were modified in accord with the biases and resolutions computed in \cite{Abbasi:2018nun} for p, He, N and Fe for the QGSJETII-04 model. For the intermediate elements we approximate these values using a $2nd$ degree polynomial interpolation. It is worth mentioning that the possible uncertainties on the bias and resolution of the intermediate elements ($\sim$ few g/cm$^{2}$), artificially introduced by this interpolation, would be much smaller than the experimental resolution of the $X_{\rm max}$ parameter (up to $20$ g/cm$^{2}$). An example of the PDFs of $X_{\rm max}$ for proton and Fe nuclei induced showers in the energy interval lg ($E$/eV) $= [18.4 - 18.5]$ as predicted by QGSJETII-04 model and modified in accord with the experimental acceptance and resolution of both experiments are displayed in Figure \ref{fig_pdfs}. Further, we will use these PDFs to fit simulated $X_{\rm max}$ distributions of random concentrations of different primaries to check the ability of the method to extract the individual fractions of nuclei.

\section{Fitting simulated $X_{\rm max}$ distributions with MC templates}

\begin{figure}
\centering
    \includegraphics[width=0.9\textwidth]{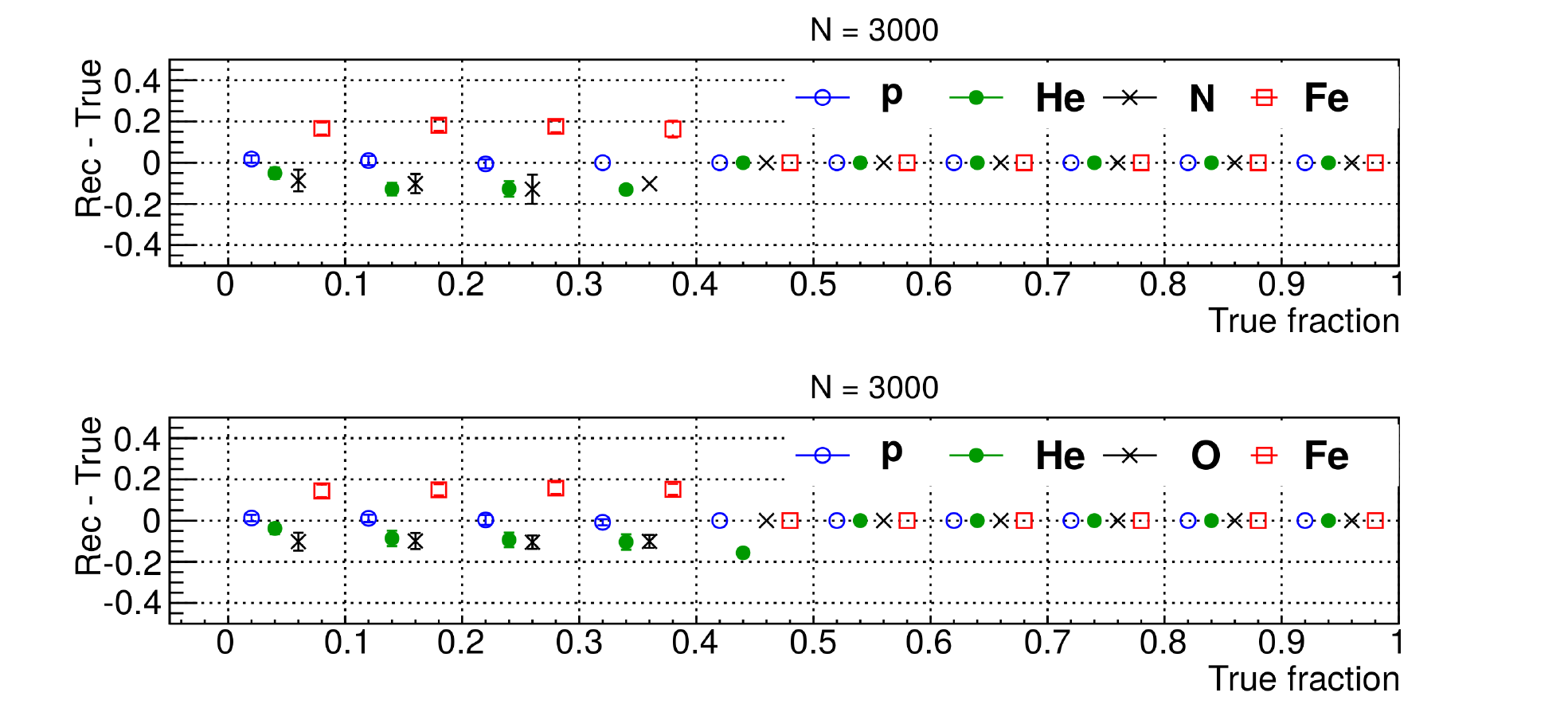}
    \caption{The bias of the reconstructed fractions used in the fitting procedure as a function of their true prior fraction, when the concentration of Si is $> 40\%$ and the $X_{\rm max}$ distributions are fitted with (p, He, N, Fe) \textit{(up)} and  (p, He, O, Fe) \textit{(down)}, in the energy interval lg ($E$/eV) $= [18.4 - 18.5]$. The statistics in $X_{\rm max}$ distribution is $N = 3000$ events. The points corresponding to the \textit{true fraction} interval $[0.4 - 1]$ can be neglected.}
    \label{true-rec}
\end{figure}

We test the ability of the method which fits the observed $X_{\rm max}$ distributions with MC templates for four primary species (p, He, N and Fe) following the binned maximum likelihood procedure and using \textit{p-value} parameter as goodness of fit. We performed the following test: we build mock data sets consisting in $X_{\rm max}$ distributions with random concentrations of 8 elements (p, He, C, N, O, Ne, Si and Fe) computed by CONEX and modified in accord with the experimental acceptance and resolution of the Pierre Auger observatory. Then we fit these mock data sets with MC templates of four fixed primary species (p, He, N and Fe) following a binned maximum likelihood procedure to extract the individual fractions of nuclei. In this fitting procedure the minimizing quantity is $-\ln L$, which is defined as:
\begin{equation}
\label{logl}
 -\ln L = \sum_{i} y_i - n_i + n_i \ln(n_i / y_i) ,
\end{equation}
where $n_i$ stands for the measured counts in the "$i$"-th bin of an $X_{\rm max}$ distribution and $y_i$ represents the MC prediction. The \textit{p-value} parameter is defined as: 
\begin{equation}
\label{pval}
\textit{p-value} = 1 - \Gamma\left(\frac{ndf}{2}, \frac{\chi^2}{2}\right), 
\end{equation}
where $\Gamma$ is the incomplete gamma function, $ndf$ represents the number of degrees of freedom, and $\chi^2$ represents the sum of the square of residuals using the parameters computed by the likelihood method.
We found that in the case in which the prior abundance of Ne or Si is quite large ($\gtrsim 40\%$) the reconstructed fractions of the four MC templates are biased from their true fractions. This effect is shown in Figure \ref{true-rec} where we represent the bias as "Rec - True" fractions as a function of prior true fraction of each element used in the fitting procedure, for the case in which the abundance of Si in the $X_{\rm max}$ distributions is grater than $40\%$ in the energy range lg ($E$/eV) $= [18.4 - 18.5]$.

\begin{figure}
\begin{minipage}[c]{0.55\linewidth}
\includegraphics[width=\linewidth]{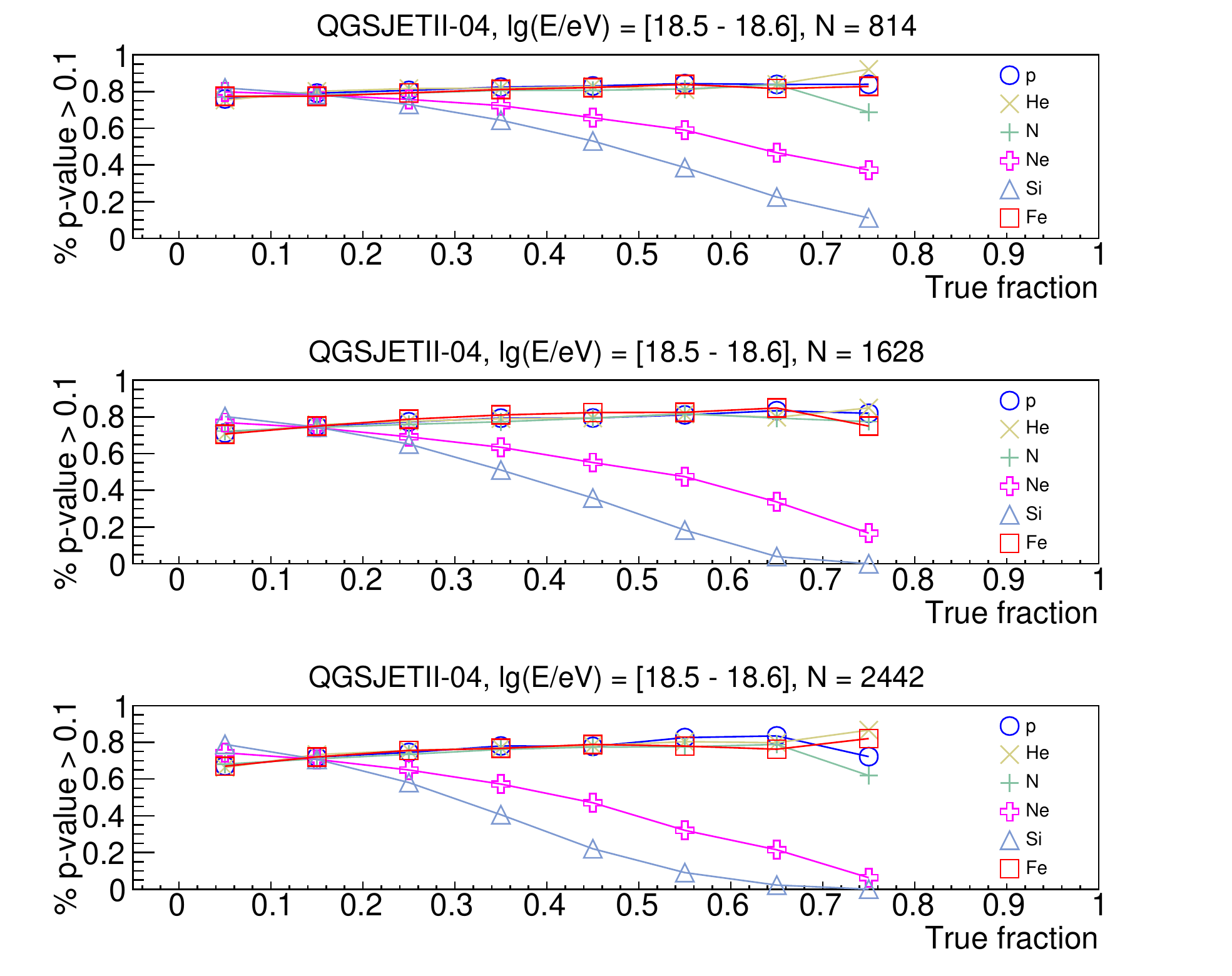}
\end{minipage}
\hfill
\begin{minipage}[c]{0.55\linewidth}
\includegraphics[width=\linewidth]{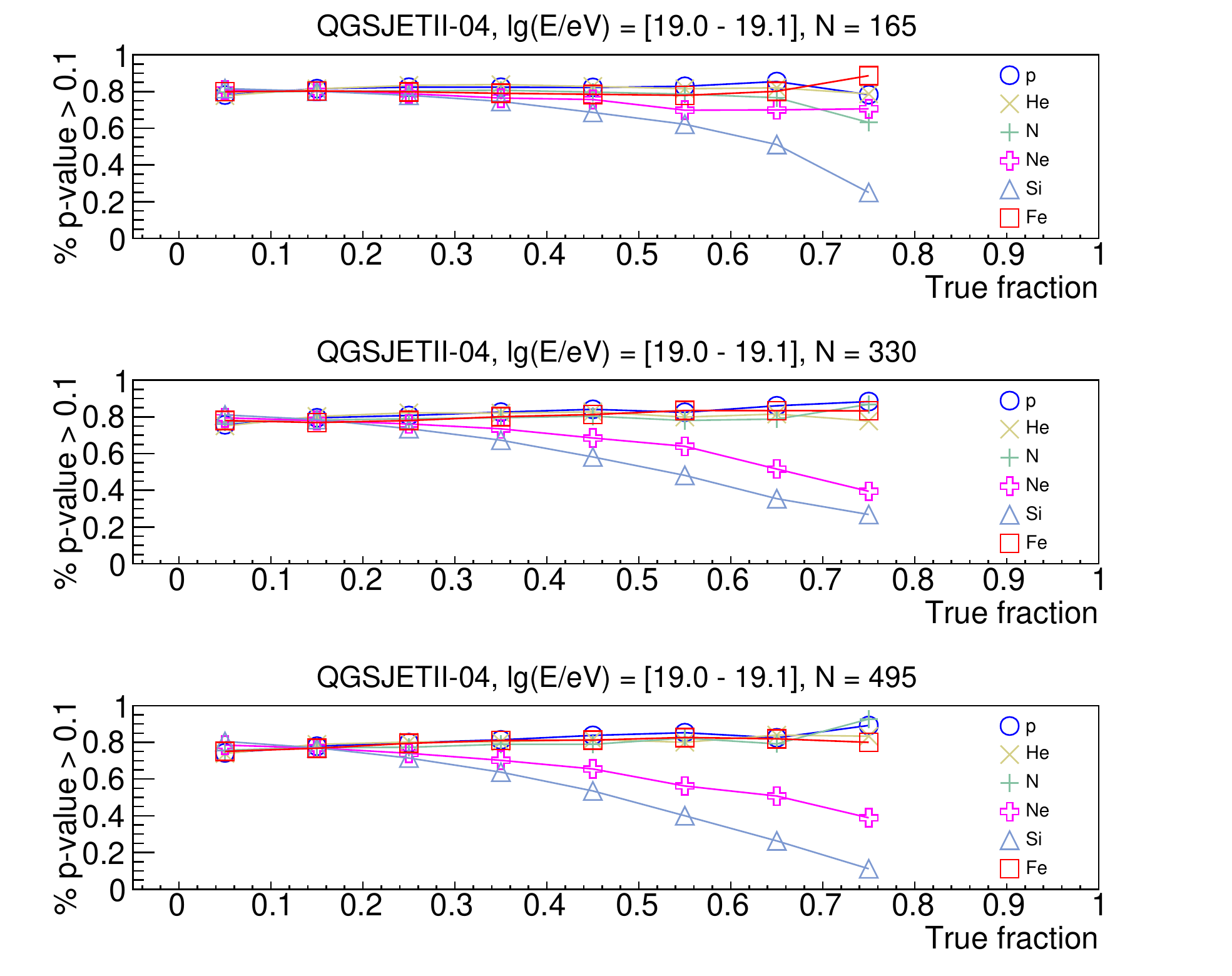}
\end{minipage}%
\caption{Fraction of events with \textit{p-value} $> 0.1$ as a function of prior abundances of different species corresponding to the energy interval  lg ($E$/eV) $= [18.5 - 18.6]$ \textit{(left)} and  lg ($E$/eV) $= [19.0 - 19.1]$  \textit{(right)}. The fitting function includes only the four fixed elements (p, He, N and Fe). The statistics of $X_{\rm max}$ distributions is indicated on the top of the plots, corresponding to the Auger statistics $N = N_{Auger}$\textit{(up)},  $N = 2 \times N_{Auger}$ \textit{(middle)} and $N = 3 \times N_{Auger}$ \textit{(down)}.}
\label{p-val}
\end{figure}

As can be seen in Figure \ref{p-val}, the probability of obtaining
a good \textit{p-value} decreases with the increase of abundances
of Ne or Si and with increase of statistics in $X_{\rm max}$ distributions, when the fitting procedure includes only four
PDFs (p, He, N and Fe). It was convenient to quantify the quality of the fit as fractions of events in which we obtained a \textit{p-value} $> 0.1$.
\section{Fitting $X_{\rm max}$ distributions recorded at Auger (2014) and TA (2016)}

\begin{figure}
\centering
    \includegraphics[width=0.8\textwidth]{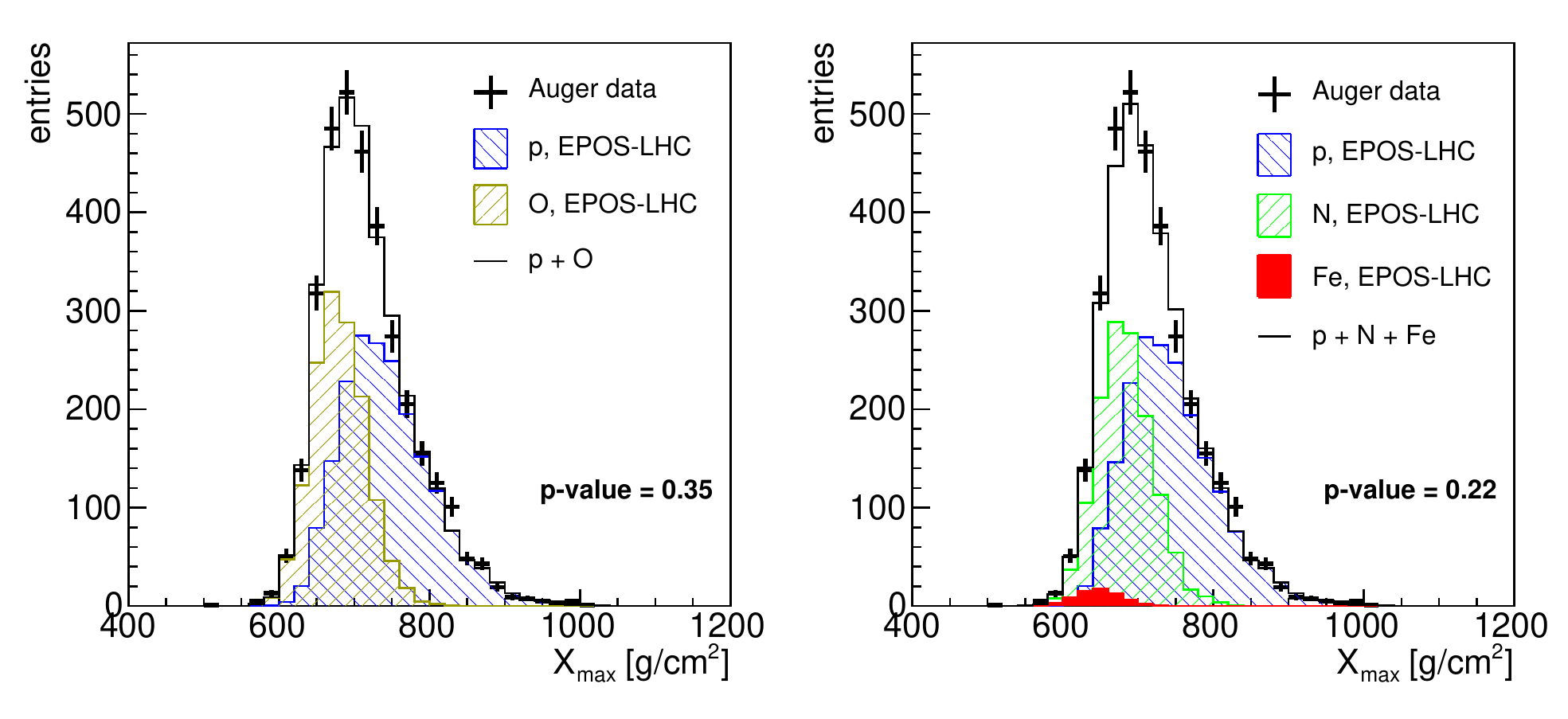}
    \caption{$X_{\rm max}$ distribution recorded by Auger in the energy range $\lg (E/\rm eV) = [17.9 - 18.0]$. The reconstructed fractions using the "best combination" approach \textit{(left)} and the method which uses the four elements (p. He, N and Fe) \textit{(right)}. Figure taken from \cite{Arsene:2020ago}.}
    \label{bestcomb}
\end{figure}

\begin{figure}
\centering
    \includegraphics[width=0.7\textwidth]{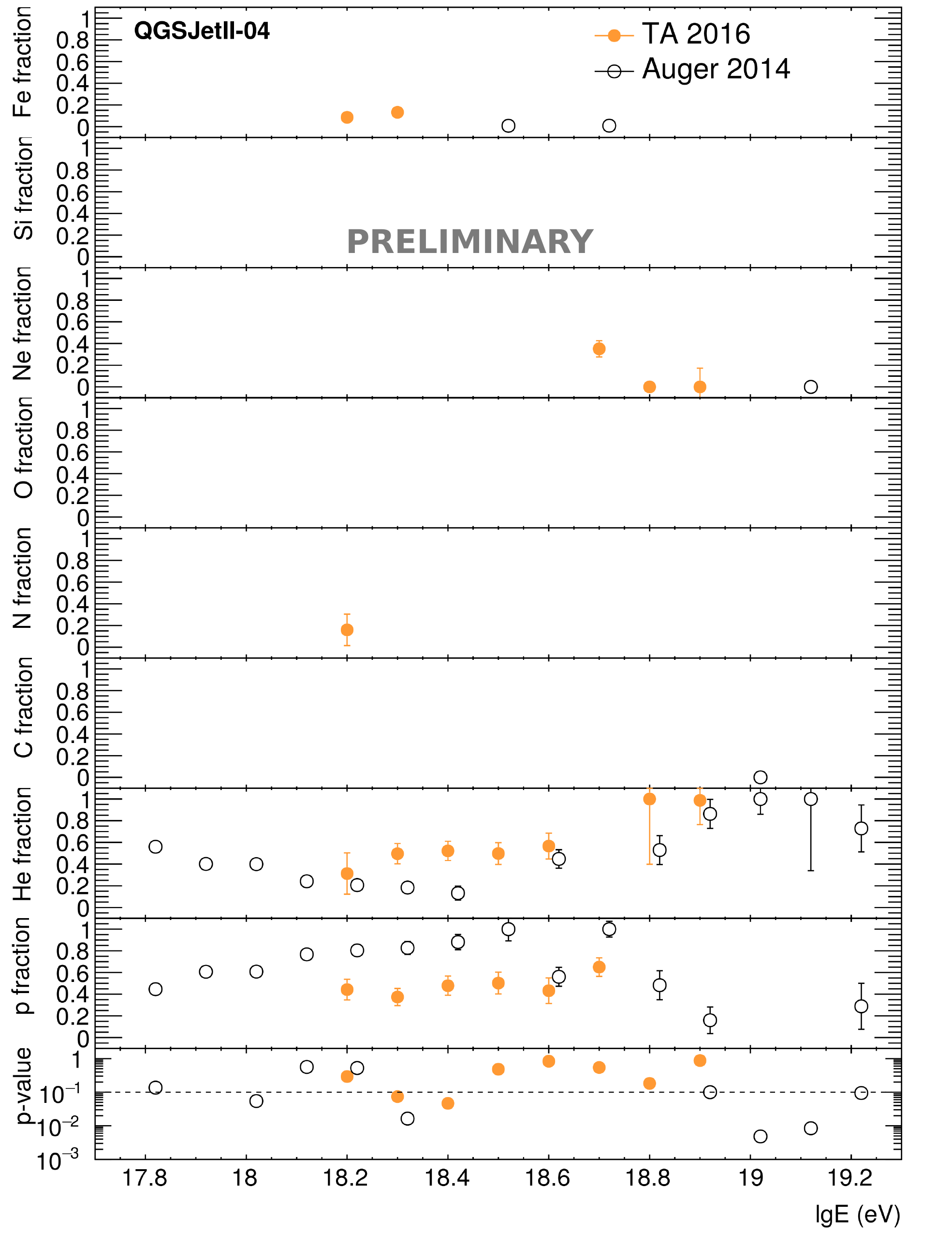}
    \caption{Fitted fractions of individual nuclei in each energy interval obtained with the "best combination" approach predicted by QGSJETII-04 model.}
    \label{auger_ta}
\end{figure}

\begin{figure}
\centering
    \includegraphics[width=0.7\textwidth]{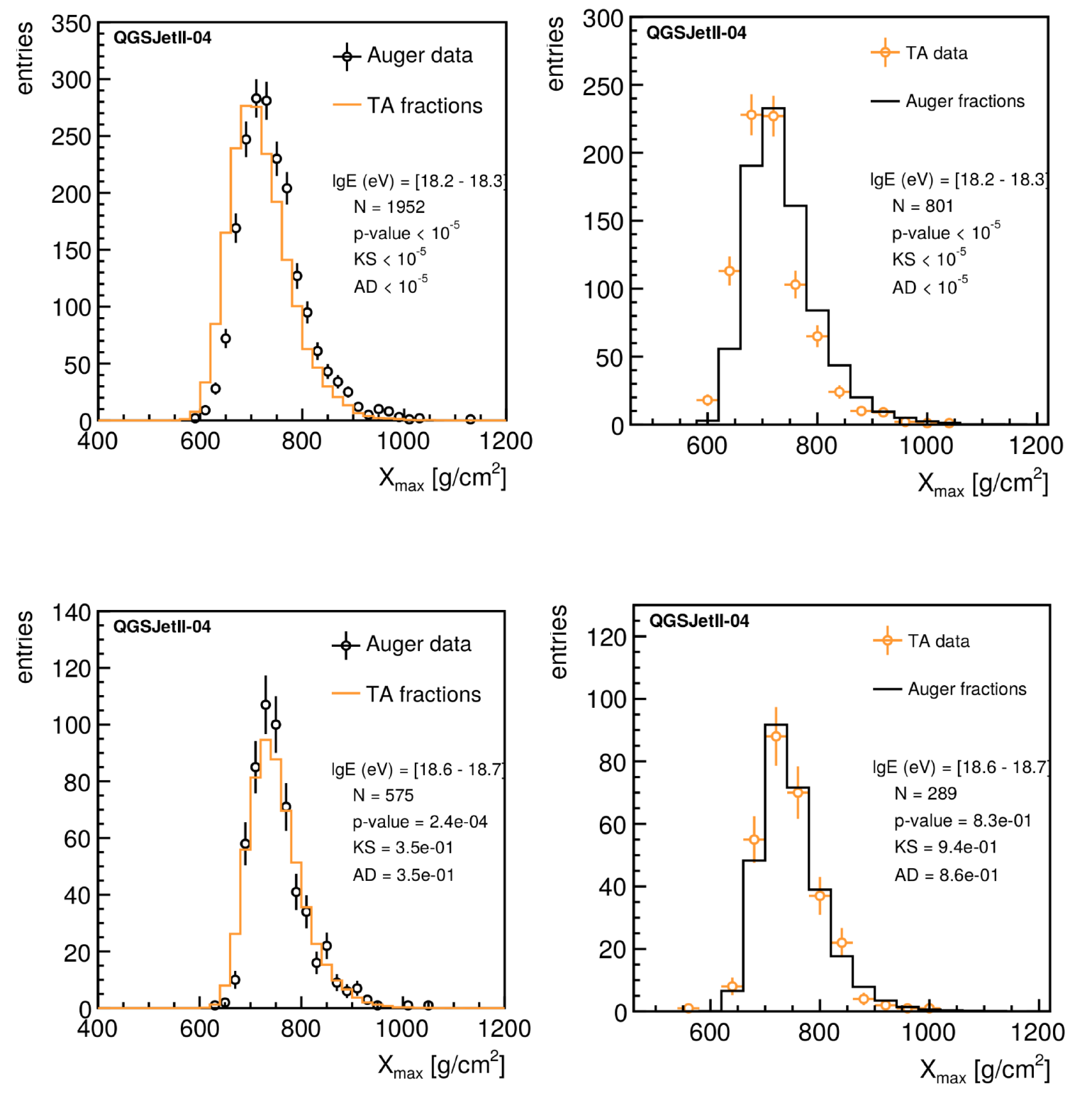}
    \caption{Comparison between Auger data and PDFs of $X_{\rm max}^{TA \rightarrow Auger}$ \textit{(left)} and TA data and  PDFs of $X_{\rm max}^{Auger \rightarrow TA}$ \textit{(right)} for the energy interval $\lg (E/\rm eV) = [18.2 - 18.3]$ \textit{(top)} and $\lg (E/\rm eV) = [18.6 - 18.7]$ \textit{(bottom)}. The parameters \textit{p-value}, $KS$ and $AD$ obtained in these comparisons are displayed on each plot.}
    \label{KS-AD-merged}
\end{figure}

We fit the experimental $X_{\rm max}$ distributions recorded at Auger (2014) and TA (2016) experiments with all possible combinations of elements from a larger set of primaries (p, He, C, N, O, Ne, Si and Fe) in each energy interval. We find the "best combination" of elements which best describe the observed distribution on the basis of the highest \textit{p-value} computet with Equation \ref{pval}. In Figure \ref{bestcomb} we give an example of an $X_{\rm max}$ distribution recorded at Auger in the energy interval $\lg (E/\rm eV) = [17.9 - 18.0]$. 
We show that the shape of the observed distribution is best described only by two elements, p and O, (Figure \ref{bestcomb} \textit{(left)}) with \textit{p-value}$= 0.35$ in comparison with the case in which we fit the experimental distribution with four elements (p, He, N and Fe) (Figure \ref{bestcomb} \textit{(right)}) with \textit{p-value}$= 0.22$.
In Figure \ref{auger_ta} we present the evolution of individual fractions of nuclei obtained with the "best combination" approach using MC templates predicted by QGSJETII-04 model.
We performed an indirect comparison between two data sets following two methods. In the first method we extract the individual fractions of each primary species reconstructed from TA data in each energy interval and build equivalent PDFs of $X_{\rm max}$ predicted for Auger (PDFs of $X_{\rm max}^{TA \rightarrow Auger}$) and the comparison is made between PDFs of $X_{\rm max}^{TA \rightarrow Auger}$ vs. Auger data while in the second method we extract the individual fractions of each primary species reconstructed from Auger data and build equivalent PDFs of $X_{\rm max}$ predicted for TA (PDFs of $X_{\rm max}^{Auger \rightarrow TA}$) and the comparison is made between PDFs of $X_{\rm max}^{Auger \rightarrow TA}$ vs. TA data. We quantify the probability of compatibility using three statistical estimators: \textit{p-value} as goodness of fit,  Kolmogorov - Smirnov (KS) and Anderson – Darling (AD). In Figure \ref{KS-AD-merged} we present such a comparison for two energy intervals $\lg (E/\rm eV) = [18.2 - 18.3]$ (containing the highest statistics $N = 1952$ events recorded by Auger) and $\lg (E/\rm eV) = [18.6 - 18.7]$ (around the \textit{ankle} with $N = 575$). As we can see, the probability of compatibility between two data sets is good in some high energy bins around and above the \textit{ankle}, but worsening with decreasing energy. The complete set of \textit{p-values}, $KS$ and $AD$ parameters obtained for each energy interval in the range $\lg (E/\rm eV) = [18.2 - 19.0]$ are displayed in Table \ref{tab}. 
\begin{table}
\caption{The probability of compatibility between two data sets as computed by \textit{p-value}, $KS$ and $AD$ tests.}
\centering
\setlength{\tabcolsep}{6pt}
\renewcommand{\arraystretch}{0.1}

\begin{tabular}{llll|lll} 
\toprule 
\multicolumn{1}{c}{} & \multicolumn{3}{c|}{Auger vs. $X_{\rm max}^{TA \rightarrow Auger}$} & \multicolumn{3}{c}{TA vs. $X_{\rm max}^{Auger \rightarrow TA}$}  \\
lgE (eV)             & p-value & KS & AD           & p-value & KS & AD          \\ 
\hline
{[}18.2 - 18.3]      &   $< 10^{-5}$           &  $< 10^{-5}$               &  $< 10^{-5}$               &	$< 10^{-5}$&	$< 10^{-5}$&	$< 10^{-5}$                                \\
{[}18.3 - 18.4]      &   $< 10^{-5}$           &  $< 10^{-5}$               &  $< 10^{-5}$               &	$< 10^{-5}$&	$< 10^{-5}$&	$< 10^{-5}$                                \\
{[}18.4 - 18.5]      &   $< 10^{-5}$           &  $2.1 \times 10^{-4}$  &  $3.6 \times 10^{-5}$  &	$< 10^{-5}$&	$3.2 \times 10^{-2}$&	$4.3 \times 10^{-3}$       \\
{[}18.5 - 18.6]      &	 $< 10^{-5}$            &  $1.1 \times 10^{-2}$  &   $2.5 \times 10^{-2}$ &	$4.0 \times 10^{-5}$&	$4.4 \times 10^{-3}$&	$2.0 \times 10^{-3}$             \\
{[}18.6 - 18.7]      &	$2.5 \times 10^{-4}$& $3.5 \times 10^{-1}$&    	$3.6 \times 10^{-1}$ &	$8.3 \times 10^{-1}$&	$9.4 \times 10^{-1}$&	$8.6 \times 10^{-1}$              \\
{[}18.7 - 18.8]      &  $< 10^{-5}$             & $6.1 \times 10^{-5}$&    	$6.3 \times 10^{-4}$ &	$4.4 \times 10^{-5}$&	$< 10^{-5}$&	$5.7 \times 10^{-4}$           \\
{[}18.8 - 18.9]      &	$< 10^{-5}$              & $< 10^{-5}$&    	$2.1 \times 10^{-4}$ &	$7.9 \times 10^{-2}$&	$7.4 \times 10^{-1}$&	$3.7 \times 10^{-1}$              \\
{[}18.9 - 19.0]      &	$7.9 \times 10^{-2}$ & $1.6 \times 10^{-2}$&    	$8.1 \times 10^{-2}$ &	$9.0 \times 10^{-1}$&	$1.0$&	$1.0$              \\
\bottomrule 
\end{tabular}
\label{tab}
\end{table}
\section{Conclusions}
In this study we present an alternative approach to infer mass composition of UHECRs from $X_{\rm max}$ distributions by fitting them with all possible combinations of elements from a large set of primaries (p, He, C, N, O, Ne, Si and Fe). 
We proved that a high prior abundance of Ne or Si ($> 40 \%$) can bias the reconstructed fractions of elements if the distributions are fitted with four fixed elements (p, He, N and Fe). We apply this method to measurements of $X_{\rm max}$ distributions recorded at Auger (2014) and TA (2016) observatories using predictions of QGSJETII-04 hadronic interaction model, concluding that the mass composition above $10^{17.8}$ eV  is dominated by protons and He nuclei on the entire energy spectrum ($\gtrsim 70\%$). An indirect comparison between the two data sets recorded by two experiments show a good degree of compatibility in some high energy intervals, especially around and above the \textit{ankle} ($\lg E (\rm eV) \sim 18.7$), but worsening at lower energies. We consider that the current approach can be used in further studies on mass composition to obtain a more accurate image of the evolution of the individual fractions of nuclei as a function of energy.

\acknowledgments{I would like to thank Octavian Sima for many useful suggestions and comments.
This work was supported by a grant of the Romanian Ministry of Education and Research, CNCS - UEFISCDI, project number PN-III-P1-1.1-PD-2019-0178, within PNCDI III.}



%
%
%

\bibliographystyle{JHEP}

\bibliography{references}

\end{document}